

\input phyzzx.tex


\def\np{Nucl. Phys.}
\def\pl{Phys. Lett.}

\def\intmp{Intern. J. Mod. Phys.}

\def\LLambda{ \tilde{ \Lambda} }


\Pubnum={SISSA/76/93/EP}
\date{June 1993}

\titlepage

\title{\bf The Complete structure of the nonlinear $W_4$ and $W_5$
algebras from quantum Miura transformation }
\author{Chuan-Jie Zhu}
\address{ International School for Advanced Studies (SISSA/ISAS)
\break {\rm and} \break INFN, Sezione di Trieste \break
via Beirut 2-4, I-34013 Trieste, Italy }

\vglue .5cm

\abstract\nobreak
{Starting from the well-known quantum Miura transformation for the Lie
algebra $A_n$, we compute explicitly the OPEs for $n=3$ and 4. The
primary fields with spin 3, 4 and 5 are found (for general $n$). By using
these primary fields and the OPEs from quantum Miura transformation, we
derive the complete structure of the nonlinear $W_4$ and $W_5$ algebras. }

\vfill\eject
\endpage

\REF\LUK{S. L. Lukyanov, Funct. Anal. Appl. {\bf 22} (1990) 1 }
\REF\ZAM{A. B. Zamolodchikov, Theor. Math. Phys. {\bf 65} (1985) 1205 }
\REF\WAA{R. Blumenhagen, M. Flohr, A. Kliem, W. Nahm,
A. Recknagel and R.  Varnhagen, \np\ {\bf B361 } (1991) 255 }
\REF\WAB{H. G. Kausch and G. M. T. Watts, \np\ {\bf B354} (1991) 740 }
\REF\MATH{S. Wolfram, {\it Mathematica, A System for Doing Mathematics
by Computer}, 2nd edition (Addison-Wesley, 1991) }
\REF\PAK{K. Thielemans, \intmp\ {\bf C2} (1991) 787 }

It is known that the quantum Miura transformation for the Lie algebra
$A_n \simeq sl(n+1) $
gives a quadratic nonlinear algebra [\LUK]. This algebra is believed to be
identical with the nonlinear extended conformal algebra $W_{n+1}$, generated
by fields $W_k$'s with the integer $k$ ranging from 2 to $n+1$. For $n=1$
and $n=2$, this
gives the Virasoro algebra and the well-known  Zamolodchikov's nonlinear $W_3$
algebra [\ZAM]. For the general case such identification is not established
explicitly. The problem with this
identification comes from the fact that the basis fields in the quantum
Miura transformation are not primary fields and the higher spin fields
in $W_{n}$ are all primary fields (by definition).
It is still an important open problem
to find a primary basis in the quantum Miura transformation.
Given the difficulty of this problem, in this paper we will establish such
identification for $W_4$ and $W_5$ (commonly known as $W_4$ and $W_5$
algebras) by explicitly computing the operator
product expansions (OPEs). In another word we will derive the complete
structure of the nonlinear $W_4$ and $W_5$ algebras directly from the
quantum Miura transformation.  In fact the structure of the $W_4$ algebra
is known in literature [\WAA,\WAB].
So our derivation serves as a non-trivial check to their results.
The method we used was then applied to derive
the more complicated $W_5$ algebra. As a first remark we note that most of
our computations are done by computer symbolic calculation
{\it Mathematica} [\MATH]. There exists also a {\it  Mathematica} package
for computing and simplifying OPEs [\PAK]\footnote{1)}{
I would like to thank A. Ganchev and De O. M. Werneck
for giving me a copy of this reference and the file of
the package.}, but I didn't make use of it in this paper.

{\bf{1. The Quantum Miura Transformation}}

\def\var{\vec{\varepsilon}}

Let $\{ \var_i, ~~i = 1, 2, \cdots, n+1\} $ be a set of vectors in an
$n$-dimensional space. They are normalized as
$$
\var_i \cdot \var_j = \delta_{ij} - { 1 \over n + 1 }, \eqn\onea
$$
and satisfy the constraint
$$
\sum_{ i = 1}^{n+1} \var_i = 0 .  \eqn\oneb
$$
Then the  quantum Miura transformation is defined as
$$
\eqalign{
R_{n+1}(z) = & :(a\partial_z + \var_1\cdot \partial_z \vec{\phi}(z) )
\cdots (a\partial_z + \var_{n+1}\cdot \partial_z \vec{\phi}(z) ) : \cr
= &\sum_{k = 0}^{n+1} U_k(z) (a\partial_z)^{n+1 - k} .
 } \eqn\onec
$$
where $:~~:$ denotes normal ordering and $a$ is a free parameter. We have
for example
$$
\eqalign{
U_0(z) = & 1, \cr
U_1(z) = & \sum_{i=1}^{n+1} \var_i \cdot \partial_z \vec{\phi} (z)  = 0, \cr
U_2(z) = & \sum_{i <j}: \var_i \cdot \partial_z \vec{\phi}(z)
 \var_j \cdot \partial_z \vec{\phi}(z) : +
a \sum_{i} (i-1) \var_i \cdot \partial_z^2 \vec{\phi}(z) .
} \eqn\oned
$$
\REF\FAT{V. A. Fateev and S. L. Lukyanov, \intmp\ {\bf A3} (1988) 507 }
For general discussion about the fields $U_k(z)$'s and their algebras, please
see refs. [\LUK, \FAT]. The fact we will explicitly verified (for $
n= 3$, 4) is that the fields $U_k(z)$'s satisfy an algebra with quadratic
defining relations
$$
U_k(z) U_l(w)  = \sum_{m\geq 2 }{ 1 \over (z-w)^m }
\sum_{ p + q = k+ l - m} C^{pq}_{kl}(m) :U_p(z) U_q(w) : + :U_k(z)U_l(w),
\eqn\onee
$$
where the coefficient $C$'s are algebraic in $a$.

{\bf 2. The fundamental OPEs}

\REF\FATA{V. A. Fateev and S. L. Lukyanov, Additional symmetries and
exactly soluable models in two-dimensional conformal field theory,''
parts I, II and III, Sov. Sci. Rev. A. Phys. {\bf 15} (1990) 1 }

To my knowledge the coefficient $C$'s are not known explicitly in general.
I suspect if ref. [\FATA] (I don't have a copy of this reference)
contains some explicit results for
these coefficients. Nevertheless ref. [\FAT] do have a general result for
$U_2(z)U_j(w)$ which is given as follows
$$ U_2(z) U_j(w) = \sum_{q=1}^j { c_q \over (z-w)^{q + 2} }  U_{j-q} (w)
+ {1  \over (z-w)^2} U_j(z)
+ { (j- 1) \over (z-w)^2} U_j(w)
+ :U_2(z)U_j(w):
, \eqn\papa
$$
where $c_q$'s are given by
$$ c_q = { (n+ 1-j+q) ! \over (n+1-j)! } \left(
j-1 + { (q-1) \over 2}  \Big( { 1\over (n+1)a^2 } + n \Big) \right) a^q .
\eqn\onecc
$$
To my knowledge the other general formula given in the same reference
for $[U_3(0), U_j(k)]$ is not sufficient to give the OPE $U_3(z)U_j(w)$.
Because of the incompleteness of these results, we will compute all the
OPEs explicitly for $n= 3$ and 4. The computation is based on the following
contraction rule
$$
\partial_z \phi_i(z) \partial_w \phi_j(w) = - { \delta_{ij} \over
(z-w)^2 } + : \partial_z \phi_i(z) \partial_w \phi_j(w):  . \eqn\onef
$$
The explicit realization of $\var_i$'s is not needed. All we needed are the
relations $\onea$ and $\oneb$. To compute the OPEs between the various
$U_k(z)$'s we first compute
$$
(a\partial_z + \var_{n+1}\cdot \partial_z\vec{\phi}(z) )U_k(w) \equiv
U_k^{n+1}(z,w) , \eqn\oneg
$$
and then
$$
(a\partial_z + \var_{n}\cdot \partial_z\vec{\phi}(z)) U_k^{n+1}(z,w) \equiv
U_k^{n}(z,w) , \eqn\oneh
$$
and etc. In each step of the above computation, there involves only one
contraction  or differentiation. This can be easily done in {\it Mathematica}.
The end result of this recursive computation gives
$$
U^1_k(z,w) = \sum_{l= 0 }^{n+1} U_l(z)U_k(w) (a\partial_z)^{n + 1 - l} .
\eqn\onei
$$
So the coefficient of the differential $(a\partial_z)^{n+1-l}$ gives the
OPE $U_l(z) U_k(w)$.

For $ n = 3$ by doing the above computation explicitly we have
$$
\eqalign{
U_2(z) U_2(w) \sim & ~~ { 3( 1 + 20 a^2) \over 2(z-w)^4 }
+ \left(  { 2\over (z-w)^2 } + { 1\over z-w} \partial_w \right) U_2(w) \cr
& + \LLambda_1(w) + (z-w) P_1(w) + (z-w)^2 \LLambda_2(w), \cr
& \cr
U_2(z) U_3(w) \sim & ~~ {6a (1 + 20a^2) \over (z-w)^5 }
+ {4a \over (z-w)^3 } U_2(w) \cr
& + \left( {3 \over (z-w)^2 } + {1\over z-w} \partial_w \right) U_3(w)
+ \LLambda_3(w) + (z-w) \LLambda_4(w) , \cr
& \cr
U_2(z) U_4(w) \sim & ~~ { 9a^2 (1+ 20a^2)\over (z-w)^6 } +
 {(1 + 36a^2) \over 4(z-w)^4 } U_2(w) +
         {3 a \over (z-w)^3} U_3(w) \cr
       & + \left( {4 \over (z-w)^2 } +
 {1\over z-w} \partial_w \right) U_4(w) + \LLambda_5(w), \cr
& \cr
U_3(z) U_3(w) \sim & - { (1+20a^2)(1 + 36a^2) \over (z-w)^6 }
- { 2 (1+12 a^2) \over (z-w)^4 } U_2(w) +{ 1  \over (z-w)^3 } P_2(w) \cr
 & -  { 1\over (z-w)^2 } \Big(  6a^2 \partial^2_w U_2(w) -
 2a \partial_w U_3(w) -
    4 U_4(w) +  \LLambda_1(w) \Big) \cr
& +  {  1 \over z-w } P_3(w)  + \LLambda_6(w), \cr
} $$
$$ \eqalign{
U_3(z) U_4(w) \sim & - { 3a (1 + 20a^2)(1 + 24a^2) \over (z-w)^7}
  - { 5a (1 + 12 a^2) \over (z-w)^5 } U_2(w) \cr
&  - { 1\over (z-w)^4 }
\Big( a(1 + 12 a^2) \partial_w U_2(w) + {3 \over 2} (1 + 8 a^2) U_3(w) \Big)
\cr
& - { 1\over (z-w)^3 } \Big(
 6 a^3 \partial_w^2 U_2(w) + { 1\over 2} \partial_w U_3(w) +  4 a U_4(w) + a
\LLambda_1(w) \Big) \cr
& + {1\over (z-w)^2 } \Big( { a \over 12} (1-24 a^2) \partial_w^3 U_2(w) + 2 a
\partial_w U_4(w) - {a\over 2} \partial_w \LLambda_1(w) - { 1\over 2}
\LLambda_3(w) \Big) \cr
& - { 1\over z-w} \Big( {1\over 2}a^3 \partial_w^4 U_2(w)
- a \partial_w^2 U_4(w) + a \LLambda_2(w) + {1\over 2} \LLambda_4(w) \Big),\cr
& \cr
& \cr
U_4(z) U_4(w) \sim & ~~ { 3 (1 + 20a^2) (5 + 180 a^2 + 2016 a^4) \over 32
 (z-w)^8} + { 5 (1 + 12 a^2)^2 \over 4 (z-w)^6 } U_2(w) + { 1\over (z-w)^5}
P_4(w) \cr
& + { 1\over (z-w)^4 } \Big( { 3\over 4 } a^2 ( 7 + 60
a^2) \partial_w^2 U_2(w) - 3 a ( 1+ 6a^2) \partial_w U_3(w) \cr
& + 3( 1 + 4a^2) U_4(w) + { 5 + 36 a^2\over 8 } \LLambda_1(w) \Big)
 + { 1\over (z-w)^3 } P_5(w)+ { 1\over z-w} P_6(w) \cr
& + { 1\over (z-w)^2 } \Big(
{ a^2 \over 16} ( 1 + 60 a^2) \partial_w^4 U_2(w) - 3a^3 \partial_w^3
U_3(w) + { 1\over 2} ( 1 + 6 a^2) \partial_w^2 U_4(w) \cr
& - { 3 \over 2} a
\partial_w \LLambda_3(w)
 - { 1\over 8} ( 1 - 36 a^2)
\LLambda_2(w) + 3 a \LLambda_4(w) + 2 \LLambda_5(w)
 - { 3\over 4} \LLambda_6(w)
\Big)  . } \eqn\onej
$$
Notice that the above formulae are not written as the form in $\onee$.
All the functions appeared in the right hand side are functions of $w$
only. Nevertheless one can explicitly verify that the above OPEs fit the
form in $\onee$. Also we have included some regular terms in
the OPEs. They are included in order to define the composite fields
$\LLambda_i(w)$'s. For the $W_4$ algebra only composite fields with spin up
to 6 are needed. The other terms denoted as $P_i(w)$ are not given explicitly.
They can easily be obtained from the symmetric property of
the OPEs: $ U_i(z) U_i(w)  = U_i(w) U_i(z)$.

For n = 4, similar OPEs between $U_k(z)$'s are also calculated but
we will not give the explicit results here because the expressions
are too long to write them. For illustration purpose we give
only the OPEs of $U_2(z)$ with other fields. These results are in agreement
with the general formula $\papa$. We have
$$
\eqalign{
U_2(z) U_2(w) \sim & ~~ { 2 ( 1 + 30 a^2) \over (z-w)^2 } +
\left( { 2 \over ( z-w)^2 } + { 1\over z-w} \partial_w \right) U_2(w) +
\LLambda_1(w), \cr
U_2(z) U_3(w) \sim & ~~ { 12 a (1 + 30 a^2) \over (z-w)^5 } + {
6 a \over (z-w)^3 } U_2(w) \cr & + \left(
 { 3 \over ( z-w)^2 } + { 1\over z-w} \partial_w \right) U_3(w) +
\LLambda_4(w), \cr
U_2(z) U_4(w) \sim & ~~ { 36 a^2 ( 1 + 30 a^2) \over (z-w)^6 } +
{ 3 ( 1 + 50 a^2)
\over 5(z-w)^4 } U_2(w) \cr &
+ { 6 a \over (z-w)^3} U_3(w) +
\left( { 4 \over ( z-w)^2 } + { 1\over z-w} \partial_w \right) U_4(w), \cr
U_2(z) U_5(w) \sim & ~~ {  48 a^3 (1 + 30 a^2) \over (z-w)^7 } +
{ 6a ( 1+ 40a^2) \over 5( z-w)^5 } U_2(w)  + { (1 + 60a^2) \over 5 (z-w)^4 }
U_3(w) \cr
& + { 4a \over (z-w)^3} U_4(w) + \left( { 5 \over ( z-w)^2 }
+ { 1\over z-w} \partial_w \right) U_5(w). \cr } \eqn\onem
$$
Here we included only two regular terms just to fix the definition of
$\LLambda_1(w)$ and $\LLambda_4(w)$. These two fields are needed to form
primary fields from $U_i(w)$'s. In the next two sections we will use these
OPEs (and the ones not explicitly written here) to derive the complete
structure of the $W_4$ and $W_5$ algebras.

4. {\bf The algebra $W_4$ }

\REF\REV{ P. Bouwknegt and K. Schoutens, Phys. Reps. {\bf 223} (1993) 183 }
\REF\NAHM{W. Nahm, in Recent developments in conformal field theories, eds.
S. Randjbar-Daemi, E. Sezgin and J. B. Zuber (World Scientific, 1990);
Conformal quantum field theories in two dimensions (World Scientific, to be
published) }

To begin with let us recall some generalities about conformal field theory.
(See ref. [\REV] for a recent review.) A primary field $\phi_h(w)$ with
dimension (spin) $h$ has the following OPE with the stress energy tensor
$T(z)$:
$$
T(z) \phi_h(w) \sim
\left( { h \over ( z-w)^2 } + { 1\over z-w} \partial_w \right) \phi_h(w),
\eqn\onen
$$
and the OPE of $T(z)$ with itself is
$$
T(z) T(w) \sim { c \over 2(z-w)^4} +
\left( { 2 \over ( z-w)^2 } + { 1\over z-w} \partial_w \right) T(w).
\eqn\oneo
$$
Here $c$ is a free parameter called center charge. From $\onej$ and
$\onem$ we can identify $U_2(z)$ with the stress energy tensor but the
fields $U_3(w)$ and $U_4(w)$ are not primary fields with spin 3 and 4.
Later we will redefine $U_i(w)$'s by adding some terms from the
descendant fields of $U_2(z)\equiv T(z)$ and other fields
such that the new fields are primary fields.
To completely fix the freedom of redefining the descendant fields, we will
ask all the fields appearing in the OPEs to be quasi-primary fields. This
is so because the OPE of two quasi-primary fields has some nice properties
[\NAHM]. The OPE of two quasi-primary fields $\phi^i$ and $\phi^j$ with
integer conformal dimensions $h_i$ and $h_j$ takes the following general
form:
$$
\phi^i(z) \phi^j(w) = { \gamma^{ij} \over (z-w)^{h_i + h_j} } +
 \sum_k C^{ij}_k \sum_{n=0}^{\infty} { a_n^{(ijk)} \over n! }
{ \partial_w^n \phi^k(w) \over (z-w)^{ h_i + h_j - h_k - n} },
\eqn\onep
$$
where $k$ denotes all the possible quasi-primary fields occurring
in the OPE (not necessarily containing singular parts), $\gamma^{ij}$
plays the role of a metric on the space of quasi-primary fields and
$ a_n^{(ijk)}$ are given by
$$ a_n^{(ijk)} = { (h_i - h_j + h_k)_n \over ( 2 h_k)_n },
\eqn\oneq
$$
with the notation $(x)_n = \Gamma(x + n) /\Gamma(x)$. Notice that for
$ h_i - h_j + h_k \leq 0$, the summation over $n$ truncates to a finite
summation.

Because of this general formula we can symbolically write the OPE of
$\phi^i(z) \phi^j(w)$ as\footnote{2)}{We use ``$\simeq$'' to signify
this writing, whereas  ``$ = $'' or ``$\sim$'' is used for other purpose.}
$$
\phi^i(z) \phi^j(w) \simeq { \gamma^{ij} \over (z-w)^{h_i + h_j} } +
 \sum_k C^{ij}_k
{ \phi^k(w) \over (z-w)^{ h_i + h_j - h_k - n} }.
\eqn\oner
$$
What this formula actually means is $\onep$. For example, the OPE
$T(z)T(w)$ expanded up to $(z-w)^2$ is
$$ \eqalign{
T(z) T(w) \sim &  2 \left(
 { 1 \over (z-w)^2 } + {1/2 \over z-w} \partial_w
 + { 3\over 20} \partial_w^2 + { 1\over 30}(z-w) \partial_w^3 +
{ 1\over 168} (z-w)^2 \partial_w^4 \right) T(w) \cr
& + \left( 1 + {1\over 2} (z-w) \partial_w +
{5\over 36} (z-w)^2 \partial_w^2\right) \Lambda_1(w) + (z-w)^2 \Lambda_2(w)+
 { c/2 \over (z-w)^4 } ,
} \eqn\ones
$$
which can be symbolically written as
$$
T(z) T(w) \simeq { c\over 2 (z-w)^4 } + { 2\over (z-w)^2 } T(w)
+ \Lambda_1(w) + (z-w)^2 \Lambda_2(w) . \eqn\onet
$$
Here the fields $\Lambda_1(w)$ and $\Lambda_2(w)$ are quasi-primary fields with
spin 4 and 6 respectively. They are related to the fields $\LLambda_1(w)$
and $\LLambda_2(w)$ in
$\onej$. One should redefine these fields such that the OPEs takes the
general form $\onep$.

As a technical remark  we mention that in $\onej$ we give only the
basic OPEs. In fact we will also need the OPEs between $U_i(z)$'s and
$\LLambda_1(w)$ and $\LLambda_1(z)$ with itself. These can be computed
easily by using the Wick theorem for the contraction involving composite
fields. See [\REV] for details.

Equipped with these general knowledges we now start our final mission. Setting
$U_2(z) \equiv T(z)$, we define new fields $W(w)$ and $U(w)$ as
follows
$$
\eqalign{
W(w) = & U_3(w) + c_1 \partial_w T(w), \cr
U(w) = & U_4(w) + c_2 \partial_w U_3(w) + c_3 \partial_w^2 T(w) + c_4
\LLambda_1(w). }  \eqn\oneu
$$
The primary field conditions are\footnote{3)}{Notice that these
formulas actually mean the following
$$
\eqalign{
T(z) W(w) \sim &
 3 \left( { 1\over (z-w)^2 } + { 1/3 \over z-w} \partial_w
 + { 1\over 14 } \partial_w^2 + { 1\over 84} (z-w) \partial_w^3 \right)
W(w) \cr
& + \Big( 1 + { 2\over 5} (z-w) \partial_w\Big)
\Lambda_3(w) + (z-w) \Lambda_4(w), \cr
T(z) U(w) \sim & 4 \left( { 1\over (z-w)^2 } + { 1/4 \over z-w} \partial_w +
{ 1\over 24} \partial_w^2 \right) U(w) + \Lambda_5(w) . } $$ }
$$
\eqalign{
T(z) W(w) \simeq & { 3\over (z-w)^2} W(w) + \Lambda_3(w) + (z-w)
\Lambda_4(w) , \cr
T(z) U(w) \simeq & { 4\over (z-w)^2 } U(w) + \Lambda_5(w),
} \eqn\onev
$$
where the fields $ \Lambda_i(w)$'s are quasi-primary. By using the OPEs
in $\onej$ we get the following unique solution for $c_i$'s
$$
\eqalign{
c_1  =&  - a, \cr
c_3 = & { (-30 + 19c + 2 c^2)\over 120 ( 22 + 5c) } , \cr}
\qquad\qquad
\eqalign{ c_2 = & - { a\over 2}, \cr
c_4 = & { (2 - 9c) \over 20 (22 + 5c) } ,
} \eqn\onew
$$
\REF\WAT{ G. M. T. Watts, \np\ {\bf B320 } (1989) 648 }
where $c$ is the central charge: $ c = 3( 1+ 20 a^2)$.
This result is in agreement with the general result for the spin-3
and 4 primary fields given in [\WAT]. Later in the next section we will
give the general formula for spin-3, 4 and 5 primary fields. The
fields $\Lambda_i(w)$ are related to $ \LLambda_i(w)$. The explicit
relations are not quite illuminating to merit displaying.

Having found the primary fields we are now ready to  compute the other three
OPEs. The computation is just a complicated algebraic calculation which can be
done by computer. The final results are
$$
\eqalign{
{ W(z) W(w) \over -( 7 +c)/10 } \simeq &
{ c/3 \over (z-w)^6 } + { 2 \over (z-w)^4 } T(w) +
{ 32 \over (22 + 5c) } { \Lambda_1(w)\over (z-w)^2 } \cr
&  - { 40 \over (7 + c) } { U(w) \over (z-w)^2}
- { 10 \over (7 + c)} \Lambda_6(w),  \cr} $$
$$\eqalign{
W(z) U(w) \simeq &  - { (c+2)(7c + 114) \over 10 (22+5c) }
{ W(w)\over (z-w)^4 }
- { 26 ( c+ 2) \over 5 (22 + 5c)} { \Lambda_3(w) \over (z-w)^2 }
- { (7c + 114) \over 10 (22 + 5c)} { \Lambda_4(w) \over (z-w) } , \cr }
$$
$$ \eqalign{
{U(z)U(w) \over { (2 +c)(7+c)(7c + 114)\over 300 (22 +5c) } }
\simeq &  { c/4\over (z-w)^8 } + { 2\over (z-w)^6 } T(w)
+ { 42 \over (22 + 5c) } { \Lambda_1(w) \over (z-w)^4 }  \cr
& + { 90 (c^2 + c + 218 ) \over (2 +c)
(7+c) (7c +114) } { U(w) \over (z-w)^4 }
 + { 3 (19c - 582)\over 10 (2+c)(7c + 114)} { \Lambda_2(w) \over (z-w)^2 }\cr
& + { 120\over (2+c)(7+c) } { \Lambda_5(w)\over (z-w)^2 }
 - { 225 (22 + 5c) \over (2 + c)(7+c)(7c + 114) }
{ \Lambda_6(w) \over (z-w)^2} \cr
& + { 96 (9c - 2) \over (2 +c)(22 + 5c)(7c +114) }
 { \Lambda_7(w) \over (z-w)^2}
, } \eqn\oney
$$
where the field $\Lambda_7(w)$ (which is quasi-primary)  is defined as
the regular part in the following OPE
$$
T(z)\Lambda_1(w) \simeq { (22 + 5c) \over 5} {T(w) \over  (z-w)^4 }
+ { 4 \over (z-w)^2 } \Lambda_1(w) + \Lambda_7(w) . \eqn\onez
$$
Up to some normalization factors for the fields $W(w)$ and $U(w)$, these
are the OPEs given in [\WAA,\WAB] for the nonlinear $W_4$ algebra.

{ \bf  5. The $W_5$ algebra }

One can repeat all the above computations for $W_5$. Before putting $n=4$
let us study the problem of finding
primary fields with spin 3, 4 and 5. To solve the primary
field condition we
need also the OPEs $U_2(z)\LLambda_1(w)$ and $U_2(z)\LLambda_4(w)$.
These OPEs are found to be $(T(z) = U_2(z))$
$$\eqalign{
T(z)\LLambda_1(w) \sim & ~~ { 3c \over (z-w)^6 } +
{ (8+c) \over (z-w)^4 } T(w) + { 3\partial_w T(w)\over (z-w)^3 }
+ \left( { 4\over (z-w)^2 }
+ { 1\over z-w } \partial_w \right) \LLambda_1(w), \cr
& \cr
T(z)\LLambda_4(w) \sim & ~~ { 7(n-1)ac \over (z-w)^7 } + { (n+1)(10+c)a\over
(z-w)^5 } T(w) + { (24 + c)\over 2(z-w)^4 } U_3(w) \cr
& + { 3\over (z-w)^3} \partial_w U_3(w) + { 2(n-1)a \over (z-w)^3}
\LLambda_1(w) + \left( { 5\over (z-w)^2 } + { 1\over z-w } \partial_w
\right) \LLambda_4(w) , } \eqn\papb$$
where $c$ is the central charge: $c = n( 1 + (n+1)(n+2) a^2 ) $.
{}From this equation and eq. $\papa$ one can prove that the following fields
are primary fields:
$$\eqalign{
W(w) = & U_3(w) - { 1\over 2} (n-1) a \partial_w T(w), \cr
U(w) = & U_4(w) - { 1\over 2} (n-2) a \partial_w U_3(w)
+ { (n-1)(n-2)\over 4n(n+1)(n+2) }
{ (2c^2 + (16+n)c - 10n ) \over (22 + 5c) } \partial_w^2 T(w) \cr
& -
{(n-1)(n-2)\over 2n(n+1)(n+2) } { ( (12 + 5n) c - 2n ) \over (22 + 5c) }
\LLambda_1(w), \cr
V(w) = & U_5(w) - { 1\over 2} (n-3) a \partial_w U_4(w) +
{ 3(n-2)(n-3) \over 4n (n+1)(n+2) } { (c^2 + (22+n)c - 18n )\over
(114+7c) } \partial_w^2 U_3(w) \cr
& - { 3(n-1)(n-2)(n-3) \over 72 n(n+1)(n+2) } { ( 2c^2 + (64 + 9n) c -
42n) a \over (114 + 7c) } \partial_w^3 T(w) \cr
& + { (n-2)(n-3) \over
n(n+1)(n+2) } { ( (20 + 7n)c -6n) \over (114 + 7c) } \Big(
{ 1\over 4} (n-1) a \partial_w \LLambda_1(w) - \LLambda_4(w) \Big) . }
\eqn\pcvv$$
Surely for $n=3$ we get back eq. $\onei$. For $n=4$ the above formulae gave
all the primary fields. Explicitly these primary fields are
$$
\eqalign{
T(w) = & U_2(w), \cr
W(w) = & U_3(w) - {  3 a\over 2} \partial_w U_2(w) , \cr
U(w) = & U_4(w) - a \partial_w U_3(w) + { (c^2 + 10c -20) \over
40 (22 + 5c) } \partial_w^2 U_2(w)
  { (4c-1) \over 5 (22 + 5c) } \LLambda_1(w), \cr
V(w) = & U_5(w) - { a\over 2} \partial_w U_4(w) +
{ (c^2 + 26c -72 ) \over 80 (114 + 7c) } \partial_w^2 U_3(w) \cr
& - { a (c^2 + 50c - 84)\over 240 (114 + 7c) } \partial_w^3 T(w) +
{ 3 a( 2c -1)\over 10 (114 + 7c) } \partial_w \LLambda_1(w)  -
{ 2 (  2c-1) \over 5 (114 + 7c) } \LLambda_4(w) .
} \eqn\twoa
$$
{}From these primary fields we can compute the $W_5$ algebra. The only
difficulty is that one should compute a lot of OPEs involving composite
fields and then introduce more composite fields until one exhausts all possible
composite fields with highest spin 8. In total there are 25 independent
composite fields which are needed in the OPEs of $W_5$\footnote{4)}{These
fields are constructed from the following normal ordered products:
$$\eqalign{
\hbox{spin 4}:&\quad T^2;\cr
\hbox{spin 5}:&\quad TW;\cr
\hbox{spin 6}:&\quad TT'',TW', TU, W^2, T^3;\cr
\hbox{spin 7}:&\quad TW'',TU',TV,WU,T^2W;\cr
\hbox{spin 8}:&\quad TT^{(4)},TW^{(3)},TU'',TV',WW'',WU',WV,
U^2,T^2T'',T^2W',T^2U,TW^2,T^4 . } $$ }. All of them are
defined to be quasi-primary. In the course of presenting the various OPEs
we will also include some regular terms in the OPEs to define some of
these composite fields. The definition of the
rest composite fields will be given after we give all
the fundamental OPEs.  Firstly the OPEs with the stress energy tensor $T(z)$
state that all the fields $W(w)$, $U(w)$ and $V(w)$ are primary fields:
$$
\eqalign{
T(z)T(w) \simeq &  { c\over 2 (z-w)^4 } + { 2\over (z-w)^2 } T(w)
+ \sum_{i=1}^3 (z-w)^{i-1} \Lambda_i(w) , \cr
T(z)W(w) \simeq & { 3\over (z-w)^2 } W(w) + \sum_{i= 0}^3 (z-w)^i \Lambda_{
i + 4} (w), \cr
T(z)U(w) \simeq & { 4\over (z-w)^2 } U(w) + \sum_{i= 0}^2 (z-w)^i \Lambda_{
i + 8} (w), \cr
T(z)V(w) \simeq & { 5\over (z-w)^2 } V(w) + \Lambda_{11}(w) + (z-w)
\Lambda_{12}(w) .
} \eqn\twob
$$
The regular terms in the above OPEs define
the composite fields $\Lambda_i(w)$ $(i=1,2,\cdots,12)$
which are all quasi-primary.
Secondly the OPEs of $W(z)$ with other fields get a little bit complicated but
still comprehensible. They are
$$
\eqalign{
{ W(z)W(w) \over c_3 } \simeq & ~~
{ c \over 3 (z-w)^6 } + { 2\over (z-w)^4 } T(w) -
{ 320 \over (68 + 7c) } { U(w) \over (z-w)^2 } \cr
& + { 32 \over (22 + 5c)} { \Lambda_1(w) \over (z-w)^2 } + \Lambda_{13}(w)
 + (z-w)^2 \Lambda_{ 14 } (w) , \cr
& \cr
W(z)U(w) \simeq & -{ 4 (2+c)(23 + c)\over 5 (22 + 5c) }{ W(w)\over (z-w)^4}
- { 208 (2+c)(23 + c) \over 5 (22 + 5c)(114 + 7c) }
{ \Lambda_4(w) \over (z-w)^2 } \cr
& +{ 5\over (z-w)^2} V(w) - { 4 (23 + c) \over 5 (22 + 5c) }
{ \Lambda_5(w)\over z-w } + \Lambda_{15} (w) + (z-w) \Lambda_{16}(w) , \cr
& \cr
W(z) V(w) \simeq & - { (116 + 3c)( 22 + 5c) \over 20 (114+7c)} {
U(w) \over (z-w)^4 } + { 3 (844 - 43c)\over 1000 (114 + 7c) }
{ \Lambda_2(w)\over (z-w)^2} \cr
& - { 2 (22+ 3c)\over (114 + 7c) } {
\Lambda_8(w)\over (z-w)^2 }  + { 3(2c-1)(68 + 7c) \over 200(114 + 7c) }
{ \Lambda_{13} (w) \over (z-w)^2 } \cr
& - { 2 (22 + 191c)\over 25(22 + 5c)
(114 + 7c) } { \Lambda_{19}(w) \over (z-w)^2 } - { 2 (116 + 3c)\over 5
(114 + 7c) } { \Lambda_9(w) \over z-w} + \Lambda_{17}(w) ,
} \eqn\opeww
$$
where $c_3 = -{ (68 + 7c)\over 80}$.
As before some regular terms are given explicitly  in order to set
the definition of the fields $\Lambda_i(w)$ $(i= 13,\cdots,17)$ which are
all quasi-primary. The definition of the spin-6 quasi-primary
fields $\Lambda_{19}(w)$ will be given later.

For the last three OPEs we will write them one by one. Firstly $U(z)U(w)$ is
given by
{ \baselineskip 10mm
$$ \eqalign{
 {U(z)U(w) \over c_4 }
\simeq & { c\over 4 (z-w)^8 } + { 2\over (z-w)^6 } T(w) +
{ 90(128 - 70c - c^2) \over (2+c)(23 + c) (68 + 7c) }{U(w) \over (z-w)^4 } \cr
& + { 42 \over (22 + 5c)} { \Lambda_1(w) \over (z-w)^4 }
+ { 1\over (z-w)^2 } \left(
 { 3 ( 19c^2 - 362 c - 7496)\over 10(2+c)(23+c)(68 + 7c) } \Lambda_2(w)
\right. \cr
&+ { 120 (118 - 7c) \over (2+c)(23 + c)(68 + 7c) } \Lambda_8(w) +
 { 9 (22 + 5c) \over 2 ( 2 + c)(23 + c) } \Lambda_{13} (w)  \cr
& \left. +
{ 72 (38 + 3c)(4c - 1)\over (2+c)(23+c)(22 + 5c)(68 + 7c) } \Lambda_{19}(w)
\right) + \Lambda_{18}(w) ,  } \eqn\opeuu
$$ }
where $c_4 = { (2+c)(23 +c)(68 +7c)\over 300 (22 + 5c) }$ and the regular term
defines the quasi-primary field $\Lambda_{18}(w)$. Secondly
$U(z)V(w)$ is
{ \baselineskip 10mm
$$
\eqalign{
U&(z)V(w) \simeq
{ (2+c)(23+c)(116+3c) \over 100 (114 + 7c)} { W(w)\over (z-w)^6}
+ { (23 + c)(116 + 3c) \over 75 (114 + 7c) } { \Lambda_5(w) \over (z-w)^3 }
\cr
& + { 1\over (z-w)^4} \left(  { 33(2+c)(23 + c)(116 + 3c) \over
50 (114 + 7c)^2 } \Lambda_4(w) + { (70272 + 9340 c + 204 c^2 + 11c^3) \over
4 (22 + 5c)(114 + 7c) } V(w) \right)  \cr
& + { 1\over (z-w)^2 } \left(
{ (7796 + 1196c + 29 c^2) \over (22+ 5c)(114 + 7c)}  \Lambda_{11}(w)
- { (334 + 37 c) \over 5 (114 + 7c) } \Lambda_{15}(w) \right. \cr
& \left. + { (2+c)(1224c^2 + 23921c - 28834)\over 25(22+5c)(114+7c)^2 }
\Lambda_{21}(w)
+ { (2+c) (297c^2 - 4934 c - 231256)\over 2100 (22 + 5c)(114+7c)
} \Lambda_6(w) \right) \cr
& +  { 1\over z-w} \left( { (13320 + 262c + 11c^2) \over 5 (22 + 5c)(114+7c) }
\Lambda_{12}(w) - { 3 (116 + 3c)\over 5 (114+7c)} \Lambda_{16}(w) \right.\cr
& \left. +
{6(2-9c)(23+c)(116+3c)\over 25(22+5c)(114+7c)^2 } \Lambda_{22}(w) +
{ 3(c-28)(23+c)(116+3c) \over 125 (114+7c)^2 } \Lambda_7(w) \right) . }
\eqn\opeuv
$$ }
Finally the last and the most complicated OPE $V(z)V(w)$ is given by
{ \baselineskip 10mm
$$
\eqalign{
{ V(z)V(w) \over c_5 } \simeq & { c\over 5 (z-w)^{10} } +
{ 2\over (z-w)^8 } T(w) + { 52 \over (22 + 5c) }
{ \Lambda_1(w) \over (z-w)^6} \cr
& + { 60 (70272 + 9340 c + 204 c^2 + 11 c^3) \over (2+c)(23+c)(68 + 7c)
(114 + 7c) } { U(w) \over (z-w)^6} \cr
& + { 1\over (z-w)^4} \left(
{ 3(1507824 + 248948c + 14880 c^2 + 181c^3)\over 2
(2+c)(23+c)(116+3c)(114+7c) } \Lambda_{13}(w) \right. \cr
& + { 24(1148c^4 + 86853 c^3 + 1942364 c^2 + 14490156 c - 3744688)\over
(2+c)(23+c)(116+3c)(22+5c)(68 + 7c)(114+7c) } \Lambda_{19}(w) \cr
& + { 1491c^4 + 55276 c^3 - 1884932 c^2 - 79552928 c - 747091776 )\over
10 (2+c)(23+c)(116+3c)(68 + 7c)(114+7c) } \Lambda_2(w) \cr
& \left. + { 120 (3767568 + 452876 c + 11520 c^2 + 187 c^3) \over
(2+c)(23+c)(116+3c)(68 + 7c)(114+7c) } \Lambda_8(w) \right)
+ { 1\over (z-w)^2} \cr
&\times  \left(
{ 4 (11c^2 - 306 c - 13656) \over
(2+c)(116 + 3c)(114+7c) } \Lambda_{14}(w)  - { 48000 \over
(2+c)(23+c)(68 + 7c)} \Lambda_{17}(w) \right. \cr
& + { 40 (609c^4 - 29492c^3 - 1718284c^2 - 49796224c - 465449792)\over
3(2+c)(23+c)(116+3c)(22+5c)(68 + 7c)(114+7c) } \Lambda_{10}(w) \cr
& + C_{20} \Lambda_{20}(w)
+ { 15360 (43 c^3 + 2393 c^2 + 23131 c - 5266) \over
(2+c)(23+c)(116+3c)(22+5c)(68 + 7c)(114+7c) } \Lambda_{23}(w) \cr
& + { 64 (114 + 7c) \over (116 + 3c)(22 + 5c) } \Lambda_{18}(w) +
{ 48 (1-2c)(578 + 19c) \over
(2+c)(23+c)(116+3c)(114+7c) } \Lambda_{24}(w) \cr
+  & \left.
 { 768 (10972 - 84704 c + 171793 c^2 + 17652 c^3 + 504 c^4 )\over
(2+c)(23+c)(116+3c)(22+5c)^2(68 + 7c)(114+7c) } \Lambda_{25}(w) + C_3
 \Lambda_{3}(w)
\right) , }\eqn\opevv
$$
where $c_5$ and the other two big coefficients are given by
$$\eqalign{
c_5 = & - { (2+c)(23 + c) (116 + 3c)(68 + 7c) \over
24000 (114 + 7c) }, \cr
C_3 = &
{  8 (5687552448 - 4443765376 c - 535589308 c^2 - 13386012 c^3 +
236551 c^4 + 4165 c^5) \over
175 (2+c)(23+c)(116+3c)(22+5c)(68 + 7c)(114+7c) } , \cr
C_{20} = &  {8 (1555590208 - 7472235776 c - 1362435108 c^2 - 56078572 c^3 +
273491 c^4 + 37380c^5 ) \over
15 (2+c)(23+c)(116+3c)(22+5c)^2(68 + 7c)(114+7c) } . } \eqn\coecc
$$ }

To finish the presentation we also need to give the definition of the
other quasi-primary fields $\Lambda_i(w)$  $(i= 19, \cdots, 25 )$. These
quasi-primary fields appear in the regular terms of the OPEs involving
quasi-primary fields. We have
$$\eqalign{
T(z)\Lambda_1(w) \simeq &
{ 22 + 5c \over 5 } { T(w) \over (z-w)^4 } + { 4\over (z-w)^2 } \Lambda_1
(w) + \Lambda_{19}(w) + (z-w)^2 \Lambda_{20}(w) , \cr
W(z) \Lambda_1(w) \simeq &
{ 48 \over 5} { W(w) \over (z-w)^4 } + { 6\over (z-w)^2 } \Lambda_4(w)
 - { 4\over z-w} \Lambda_5(w)\cr
&  + \Big(
\Lambda_{21}(w) + { 64 \over 35} \Lambda_6(w) \Big) + (z-w) \Lambda_{22}(w) ,
\cr
T(z)\Lambda_8(w) \simeq & { 24 + c\over 2 (z-w)^ 4 } U(w) +  { 6\over (z-w)^2}
\Lambda_8(w) + \Lambda_{23}(w) , \cr
T(z)\Lambda_{13}(w) \simeq & {46\over 63} { T(w) \over (z-w)^6 }
- { 3520 \over 3 (68 + 7c) } { U(w)\over (z-w)^4 }
+  { 6\over (z-w)^2}
\Lambda_{13} (w) + \Lambda_{24}(w) , \cr
T(z)\Lambda_{19}(w) \simeq &  {8(22+5c)\over 15} { T(w) \over (z-w)^6 }
+  { 6\over (z-w)^2}
\Lambda_{19} (w) + \Lambda_{25}(w) . \cr
}\eqn\defsu
$$
Of course, these OPEs involving quasi-primary fields aren't all the OPEs
needed in the derivation of the $W_5$ algebras.

{\bf 6. Discussion }

{}From our explicit computation we see that there always
exists a unique  spin-$j$
primary field for $j=3$, 4 and 5. The  general formula is given by
eq. $\pcvv$.
Here we explain why this is so. For $U_3(w)$ there are two
anomalous terms (the central term $1/(z-w)^5$ and $T(w)/(z-w)^3$)
to be cancelled but we only have one freedom by adding
$\partial_w U_2(w)$ to $U_3(w)$. Nevertheless these two anomalous terms are
related as shown by the following general observation: the OPE
$(j > 2)$
$$ T(z)W_j(w) \sim { c_j\over (z-w)^{j+2} } +
\left( {  j \over (z-w)^2 } + { 1\over z-w} \partial_w \right) W_j(w) ,
\eqn\ppa $$
satisfies the Jacobi identity:
$$
[  [ L_m, L_n ], W_j(p)] + \big(
[ [ W_j(p), L_m] , L_n] - (m \leftrightarrow n) \big) = 0. \eqn\ppcd
$$
only for $c_j = 0 $. So the vanishing of one anomalous term insures the
vanishing of the other term because the original algebra surely satisfies the
Jacobi identities and redefinition of fields doesn't spoil this property.
For $U_4(w)$ there are four anomalous terms but only three of them are
independent.
Here we have three terms: $\partial_w U_3(w)$, $\partial_w^2 U_2(w)$ and
$\LLambda_1(w)$ to be added to $U_4(w)$. So a unique spin-4 primary field
always exists. For higher spin fields a simple counting of freedoms can't
prove the existence and uniqueness of higher spin  primary fields.

What we can learn about the general structure of nonlinear $W$-algebras
from our explicit results?
By looking at the explicit OPEs we can  guess  about the first few
terms in $W_j(z) W_j(w)$ as the following
$$
W_j(z) W_j(w) \simeq { c/j \over (z-w)^{2j} }
+  { 2\over (z-w)^{2(j-1)} }  T(w)
+ { 2(5j+1)\over (22 + 5c) }{ \Lambda_1(w) \over (z-w)^{2(j-2)} }  + \cdots .
\eqn\conm  $$
where $W_j(w)$ is a spin-$j$ $(j>2)$ primary field.
There is nothing special about the first term because it just set the
normalization for $W_j(w)$. The second term can be proved by considering
the central term in the following Jacobi identity
$$
[  [ W_j(m), W_j(n)] , L_p] + \big(
[ [ L_p, W_j(m)] , W_j(n)] - (m \leftrightarrow n) \big) = 0. \eqn\wwt
$$
The other term in $\conm$ is an extrapolation from our explicit results.
It is known to be true in $W(2,\delta)$ algebras [\WAA]. A proof could be
found by extending (or just following) their computations.

As a further remark we notice that these three terms in $W_j(z)W_j(w)$ are
content independent, meaning that the structure constants only depend
on the spin of the primary field $W_j(w)$ and don't depend on the content
of the algebra, i.e. how many basic fields constitute the algebra. We
conjecture that these three structure constants are the only content
independent ones. This conjecture is supported by our explicit results for
the nonlinear $W_4$ and $W_5$ algebras. It is also true
in $WB_2$, a nonlinear extended conformal algebra with a spin-4 primary
field which is associated with the simple group $B_2$ or $C_2$.

\REF\ADDA{A. Konecker, Automorphisms of W-algebras and extended rational
conformal field theories, preprint BONN-HE-92-37 (November 1992) to
be published in Nucl. Phys. B }
\REF\ADDB{K. Hornfeck, \pl\ {\bf B275} (1992) 355 }
\REF\ABBC{K. Hornfeck, W-algebras with set of primary fields of dimensions
$(3,4,5)$ and $(3,4,5,6)$, preprint KCL-TH-9209 or DFTT-70/92
(December 1992) }

One other aspect of the OPEs of the $W_n$ algebras is that there is a
selection rule. This is closely related to the automorphisms of $W$-algebras
which was discussed extensively in ref. [\ADDA]. The primary fields fall
into two sets: the even set consisting
of even spin primary fields and the odd set consisting of
odd spin primary fields.
The OPEs of (even)$\times$(even) and (odd)$\times$(odd) fields give
only even fields and the OPEs of (even)$\times$(odd) fields give only
odd fields. Presumably this selection rule is also presented in
$W_n$-algebras [\ADDA].

\vglue .5cm

The author would like to thank Prof. R. Iengo and Dr. D. P. Li
for interesting discussions.
This work at SISSA/ISAS was supported by an INFN post-doctoral
fellowship.

\vglue .5cm

Note added: After I finish this paper, I became aware of two papers by
K. Hornfeck [\ADDB, \ABBC] which also studied the $W_5$ algebra by
using Jacobi identities. In paper [\ABBC], some structure constants
are also computed from quantum Miura transformation.

\vfill\eject
\refout
\bye